\documentclass[conference]{IEEEtran}
\usepackage{ifpdf}

\usepackage{cite}

\ifCLASSINFOpdf
  \usepackage[pdftex]{graphicx}
\else
  \usepackage[dvips]{graphicx}
\fi
\usepackage{amsmath, amssymb}
\usepackage{algorithm, algorithmic}

\usepackage{array}
\usepackage{fixltx2e}

\usepackage{stfloats}
\usepackage{url}

\usepackage[colorlinks,
linkcolor=red,
anchorcolor=green,
citecolor=blue]{hyperref}

\usepackage{multirow}
\usepackage{booktabs}
\usepackage{arydshln}
\usepackage[table,xcdraw]{xcolor}
\usepackage{longtable}
\usepackage{booktabs}
\usepackage{threeparttable}
\usepackage{lscape}
\usepackage{subfigure}
\usepackage{threeparttable}
\usepackage{verbatim}
\usepackage{marvosym}
\usepackage{ulem}
\usepackage{fancyhdr}
\usepackage{soul}
\usepackage{pifont}

 \newcommand{\rOneFC}[1]{\textcolor{black}{#1}}
\newcommand{\rTwoFC}[1]{\textcolor{black}{#1}}
\newcommand{\rOneCN}[1]{\textcolor{black}{#1}}
\newcommand{\rTwoCN}[1]{\textcolor{black}{#1}}
\newcommand{\rThreeCN}[1]{\textcolor{black}{#1}}

\newcommand{\Rone}[1]{\textcolor{black}{#1}}
\newcommand{\rnewOneCN}[1]{\textcolor{black}{#1}}
\newcommand{\rfinalCN}[1]{\textcolor{black}{#1}}
\newcommand{\WX}[1]{\textcolor{black}{#1}}
\newcommand{\FC}[1]{\textcolor{black}{#1}}
\newcommand\Mark[1]{\textsuperscript#1}

\begin{document}

\title{\rTwoCN{A Scalable RISC-V Vector Processor \\ Enabling Efficient Multi-Precision DNN Inference} }

\vspace{-6em}
\author{
    Chuanning~Wang\Mark{1}, Chao~Fang\Mark{1}, Xiao~Wu\Mark{1}, Zhongfeng~Wang\Mark{1}$^{,}$\Mark{2}$^{(\textrm{\Letter})}$, Jun~Lin\Mark{1}$^{(\textrm{\Letter})}$ \\
	\IEEEauthorblockA{
		\Mark{1}School of Electronic Science and Engineering, 
		Nanjing University, Nanjing, China\\
            \Mark{2}School of Integrated Circuits, Sun Yat-sen University, Shenzhen, China\\
		Email:
        \{chuan\_nwang, fantasysee, wxiao\}@smail.nju.edu.cn, \{zfwang, jlin\}@nju.edu.cn
    }
}
\maketitle
\vspace{-6em}

%


\pagestyle{fancy}
\lhead{\footnotesize Accepted by 2024 IEEE International Symposium on Circuits and Systems (ISCAS 2024)}
\renewcommand\headrulewidth{0pt}




\begin{abstract}
RISC-V processors encounter substantial challenges in deploying multi-precision deep neural networks (DNNs) due to their restricted precision support, constrained throughput, and suboptimal dataflow design.
\rTwoFC{To tackle these challenges, a scalable RISC-V vector (RVV) processor, namely SPEED, is proposed to enable efficient multi-precision DNN inference by innovations from customized instructions, hardware architecture, and dataflow mapping.}
\rTwoFC{Firstly, dedicated customized RISC-V instructions are proposed based on RVV extensions, providing SPEED with fine-grained control over processing precision ranging from 4 to 16 bits.}
\rnewOneCN{Secondly, a parameterized multi-precision systolic array unit is incorporated within the scalable module to enhance parallel processing capability and data reuse opportunities.}
\rnewOneCN{Finally, a mixed multi-precision dataflow strategy, compatible with different convolution kernels and data precision, is proposed to effectively improve data utilization and computational efficiency.}
We perform synthesis of SPEED in TSMC 28nm technology.
\rnewOneCN{
The experimental results demonstrate that SPEED achieves a peak throughput of 287.41 GOPS and an energy efficiency of 1335.79 GOPS/W at 4-bit precision condition, respectively.
Moreover, when compared to the pioneer open-source vector processor Ara, SPEED provides an area efficiency improvement of 2.04$\times$ and 1.63$\times$ under 16-bit and 8-bit precision conditions, respectively, 
which shows SPEED's significant potential for efficient multi-precision DNN inference.}
\end{abstract}
\section{Introduction} \label{sec:intro}
\rOneFC{RISC-V processors \cite{xu2022towards, Rossi2021VegaAT, Garofalo2023Dustin, Garofalo2023DARKSIDEAH, yun2023TCSII, Cavalcante2022NewAra, Cavalcante2020AraA1TVLSI, Askarihemmat2023QuarkAI, Theo2023sparq, he2023agile, huang2023precision} are distinguished by the concise open-source RISC-V instruction set architecture (ISA) \cite{Waterma2016EECS}, which defines a fundamental set of instructions while offering opportunities to incorporate application-specific custom instructions.}
\rOneFC{This unique feature enables RISC-V processors as a promising solution ranging from energy-efficient embedded systems \cite{he2023agile} to high-throughput servers \cite{xu2022towards}.}
\rnewOneCN{On these RISC-V hardware platforms, deep neural networks (DNNs) exhibit substantial deployment demands~\cite{askarihemmat2023barvinn, wu2021flexible, fang2021accelerating} but face the extensive storage and computational costs. 
Aiming to reduce resource overhead and improve inference speed, multi-precision quantized DNNs~\cite{ding2018quantized, tian2023bebert, zhou2018explicit} have emerged as an efficient choice for deployment.}
\rOneFC{However, deploying multi-precision quantized DNNs also incurs a series of significant challenges on prior RISC-V processors.}

{The emergence of the RISC-V Vector (RVV) extension instruction set has made RVV processors a promising choice for deploying multi-precision DNNs.}
\rOneCN{These RVV processors \cite{Cavalcante2022NewAra, Cavalcante2020AraA1TVLSI, Askarihemmat2023QuarkAI, Theo2023sparq, yun2023TCSII} excel in throughput enhancement through their parallel processing capabilities and minimize instruction overhead by employing a \rnewOneCN{minimal} number of configuration-setting instructions to define data precision.}
\rTwoFC{Notably, Ara~\cite{Cavalcante2022NewAra} stands out as the pioneer open-source vector processor compatible with the standard RVV 1.0 ISA. It achieves an exceptional throughput improvement of up to 380$\times$ in comparison to a scalar core~\cite{zaruba2019cost}, rendering it highly efficient for tasks such as DNN inference.}
However, the potential performance improvement of the Ara is constrained due to limited support for low-bitwidth (e.g., 4-bit) operations,
which are widely used in quantized DNNs to reduce computational complexity and accelerate inference processing with little loss in accuracy~\cite{Li2022APE}.
\rTwoCN{Furthermore, Ara's parallelism is constrained by the number of scalable modules.}
When handling intensive computational tasks, increasing the number of scalable modules can cause excessive hardware consumption.
\rnewOneCN{Meanwhile, inefficient dataflow of Ara can lead to increased off-chip data movement and underutilization of on-chip memory, thereby lowering computational efficiency.}
\rTwoFC{In summary, deploying multi-precision quantized DNNs on prior RISC-V processors still struggles with (1) limited precision support, (2) constrained throughput, and (3) inefficient dataflow mapping.}

\rThreeCN{To address the issues above, we propose SPEED, a scalable RVV processor enabling efficient DNN inference across 4, 8, and 16 bits through customized instructions based on the standard RVV ISA.}
\rnewOneCN{Moreover, SPEED significantly improves data throughput and computational efficiency by increasing the data-level parallelism of scalable modules and utilizing a mixed multi-precision dataflow strategy tailored for various convolution kernels.}
Specifically, the contributions of this work can be summarized in the three following aspects:

\begin{enumerate}
    \item[\textbf{1)}] \textbf{Customized instructions}: 
    Dedicated customized RISC-V instructions are proposed based on RVV extensions, providing SPEED with fine-grained control over processing precision and dataflow strategy, \rnewOneCN{while supporting multi-precision computations ranging from 4 to 16 bits.}
    \item[\textbf{2)}] \textbf{Hardware architecture}: 
    \rTwoFC{A scalable RVV processor is developed, namely SPEED, for efficient multi-precision DNN inference by enhancing parallel processing capability of scalable modules.}
    Compared to Ara\cite{Cavalcante2022NewAra}, 
    SPEED improves an area efficiency by 2.04$\times$ and 1.63$\times$ under 16-bit and 8-bit precision conditions, respectively.
    \item[\textbf{3)}] \textbf{Dataflow mapping}: 
    A mixed dataflow strategy embracing feature map-first (FF) and channel-first (CF) strategies is proposed,
    aiming to achieve high computational efficiency of convolution kernels across various kernel sizes and data precisions.
    The mixed dataflow strategy improves area efficiency of SPEED by 1.87$\sim$3.53$\times$ over Ara \cite{Cavalcante2022NewAra} under various convolution kernels.
\end{enumerate}
\vspace{-1em}

\begin{figure}[!htbp]
  \centering
    \includegraphics[width=1\columnwidth]{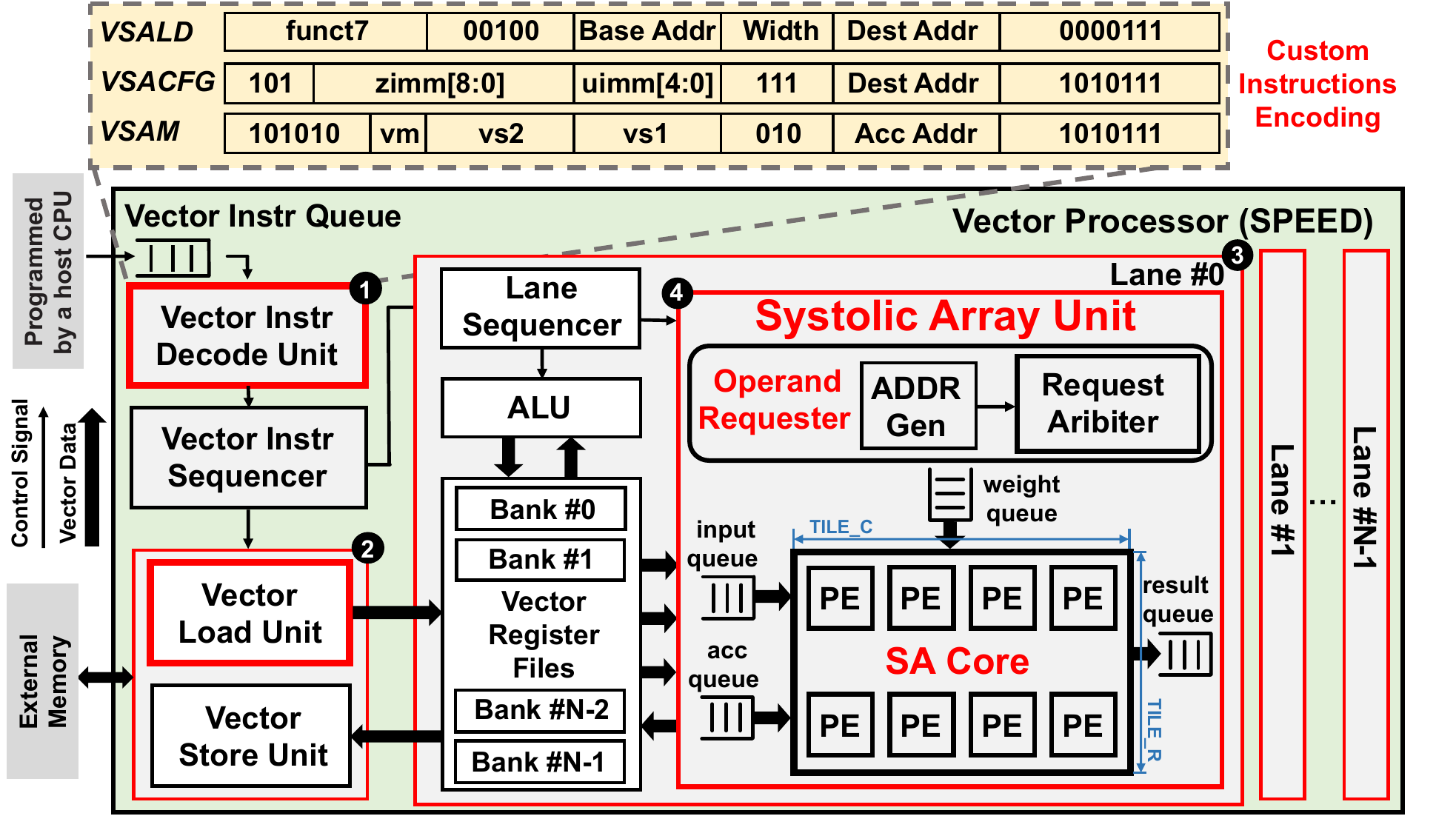}
      \caption{Customized instructions and overall architecture of SPEED.}
        \label{fig:full_architecture}
   \vspace{-1.8em}
\end{figure}

\section{The Proposed SPEED Architecture}\label{sec：design}
 
SPEED is a scalable RVV processor augmented with customized instructions enabling high-throughput multi-precision DNN inference, which is built on the RVV v1.0 ISA. 
Additionally, the RVV processor is tightly coupled to a RISC-V scalar core for programmable instructions and an external memory for fetching necessary data.
In this section, SPEED will be described from three perspectives: customized instructions, hardware architecture, and dataflow mapping.
\vspace{-0.5em}
\subsection{Customized Instructions} \label{subsec:customized_ins}
To facilitate efficient inference of multi-precision DNNs, several customized instructions are proposed to enhance the computational efficiency of DNN operations compared to the standard RVV extensions.
Specifically, the customized instructions mainly contain the configuration-setting (\texttt{VSACFG}), load (\texttt{VSALD}), and arithmetic (\texttt{VSAM}), as depicted in \Rone{Fig.~\ref{fig:full_architecture}}.

\texttt{VSACFG} serves as a vector configuration-setting instruction that effectively provides the necessary \rfinalCN{information for subsequent instructions, such as data precision (4-$\sim$16-bit) and dataflow strategy (FF/CF strategy).}
This information is encoded within the \textit{zimm9} and \textit{uimm5} encoding spaces, as shown in \Rone{Fig.~\ref{fig:full_architecture}}, 
\rnewOneCN{to prepare for processing of subsequent RVV and customized instructions.}

\texttt{VSALD} is a customized load instruction responsible for loading data from the base address of external memory and storing it into vector register files (VRFs) at the specified destination address, aiming to maximize data utilization.
In contrast to the ordered allocation operation of the standard RVV load instruction \texttt{VLE}, the loaded data from the external memory are broadcast to each lane for improving data reuse.
 
\rnewOneCN{\texttt{VSAM} is a customized arithmetic instruction that exploits data-level parallelism in efficiency,}
thereby enhancing computational efficiency.
As shown in Fig.~\ref{fig:full_architecture}, \texttt{VSAM} requests data from the base addresses \textit{vs1} and \textit{vs2} where they locate in the VRFs, respectively. 
The performed results of above data are stored at the accumulation address \textit{Acc Addr} within the VRFs.

\subsection{Hardware Architecture}\label{subsec:overall_archi}
The overall hardware architecture of SPEED is shown in Fig.~\ref{fig:full_architecture}.
To realize efficient multi-precision DNN inference, 
\ding{202}~vector instruction decode unit (VIDU) is developed to decode customized instructions as well as the standard RVV instruction set.
\rnewOneCN{Furthermore, \ding{203}~vector load unit (VLDU) is designed to distribute data through broadcast or ordered allocation,}
enabling our design to meet the diverse computation requirements of mixed dataflow strategy.
Scalable modules for vector processors, namely \ding{204}~lane, serve as the main computational components of the proposed processor, which consists of lane sequencer, VRFs, systolic array unit (SAU) and arithmetic logic unit (ALU).

To enhance the processor's parallel processing capability and \WX{fully exploit data reuse opportunities,}
\WX{a highly flexible and parameterized multi-precision \ding{205}~SAU is presented as the main computing unit of SPEED, which is composed of three components: operand requester, queues, and systolic array core (SA Core).}
\WX{The operand requester consists of an address generator and a request arbiter, enabling efficient data access by concurrently generating addresses and prioritizing data requests.}
The queue is responsible for buffering the data involved in the computation, including inputs, weights, accumulation results, and outputs.
The SA Core is a reconfigurable two-dimensional array of processing elements (PEs) determined by the parameters $TILE\_R$ and $TILE\_C$, which can be flexibly adjusted for different computation requirements.
For convolution operations, SA Core architecture employs three levels of parallelism: within each PE on input channel dimension, across the PE array within each lane on output channel dimension and on the height dimension of feature map. 
\rfinalCN{Moreover, each PE consists of sixteen 4-bit multipliers that can be dynamically combined to perform  multiply-accumulate operation (MAC) with 16-bit precision, four sets of MACs at 8-bit precision, or sixteen sets of MACs at 4-bit precision.
}
\subsection{Dataflow Mapping}
\label{subsec:mix_flow}
To enhance SPEED's computational efficiency, an optimized multi-precision dataflow strategy, combined with FF and CF strategies, is developed to flexibly handle DNN layers with varying kernel sizes.
Specifically, the proposed FF strategy is suitable for convolutional layers with large kernel size, while the CF strategy is better suited for those with small kernel size.
Detailed descriptions are provided as follows.

\begin{figure}[t]
    \centering
    \includegraphics[width=\columnwidth]{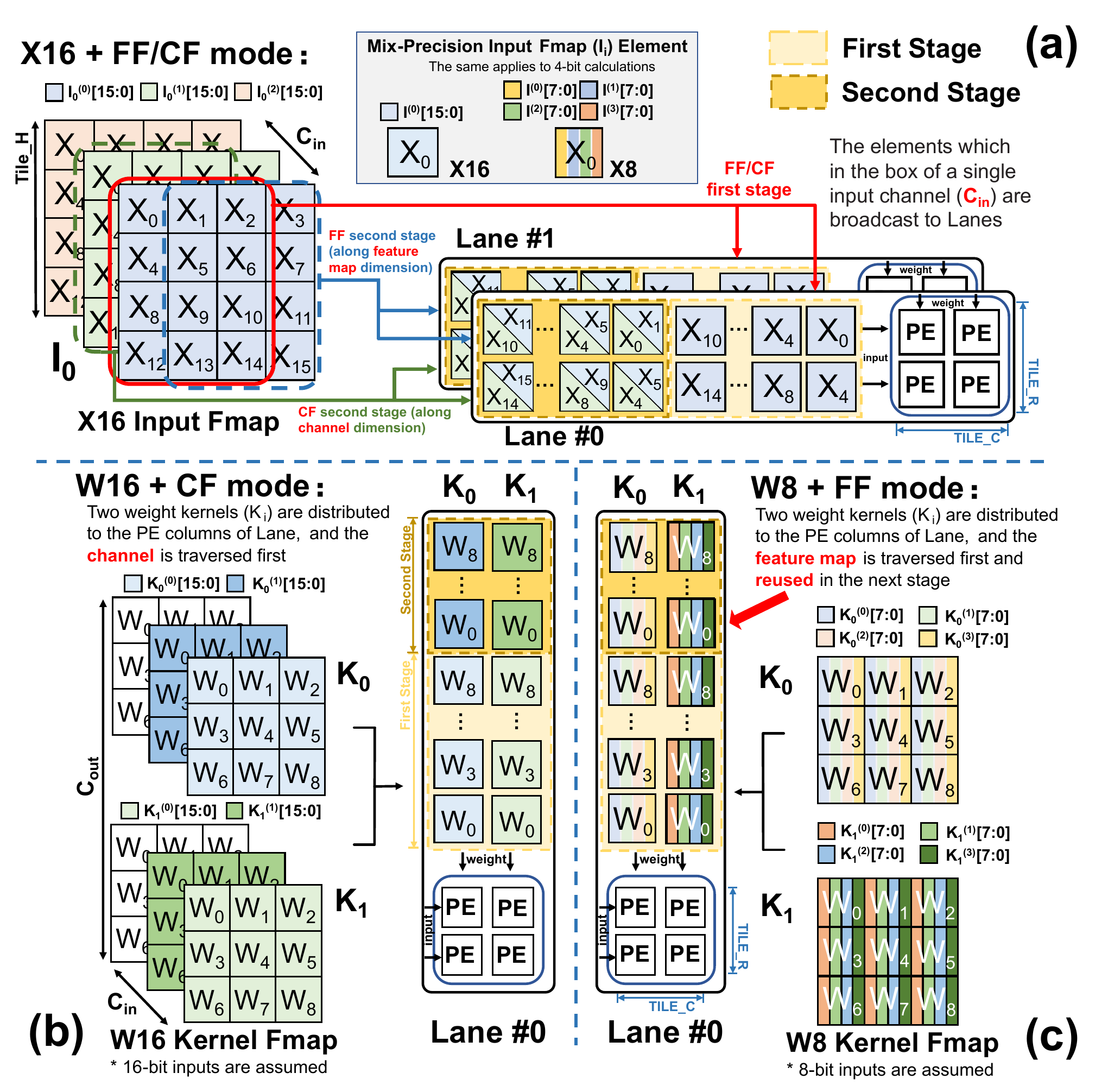}
    \caption{Examples on how CF strategy and FF strategy work with multi-precision elements. Note that the 4-bit data is operated in the same way as 16-bit and 8-bit.}
    \vspace{-1.2em}
\label{fig:mult_precision_FF_CF}
\end{figure}

\FC{To unify multi-precision data representation, all inputs and weights are preprocessed into multiple elements along the input channel dimension. Specifically, every adjacent 1, 4, and 16 operands are combined into a unified element under 16-bit, 8-bit, and 4-bit precision modes, respectively.}
\rnewOneCN{Both the FF and CF strategies support continuous multi-stage convolution computations.}
\rnewOneCN{To clearly show how these strategies work, we select the first and second stages as examples of broadcasting the inputs and allocating the weights to the lanes, as shown in \Rone{Fig.~\ref{fig:mult_precision_FF_CF}}.}
\rnewOneCN{Here, the size of weight kernel is $3\times 3$, and the parameter $TILE\_H$ is set to 4, which is determined by the parallelism of the SAU and kernel size.}

For example, as shown on the left of \Rone{Fig.~\ref{fig:mult_precision_FF_CF}~(a)}, 16-bit input (X16) elements are pre-fetched to the VRF during two-stage convolution operations.
\rnewOneCN{In FF strategy, 
to facilitate parallel computing and reduce off-chip data movement, 
we pre-fetch $4 \times 4$ elements on a single input channel of inputs.}
\rnewOneCN{These pre-fetched elements are requested from the VRFs and calculated in the SAU during the two-stage convolution operations, 
and the request process is as depicted in the right of \Rone{Fig.~\ref{fig:mult_precision_FF_CF}~(a)}.
When the first stage is completed, the results are stored to the VRFs, 
and the elements in the overlapping areas of the blue and red boxes are reused.}
\rnewOneCN{However, multi-stage convolution results occupy a large portion of the storage space in VRFs, and extra time is wasted in transferring the partial results between stages.}
\rnewOneCN{In order to reduce the memory footprint of VRFs and avoid output transfer latency, 
the CF strategy is proposed to pre-fetch elements along the input channel dimension.}
\rnewOneCN{The two-stage convolution operations pre-fetch elements from the two input channels of inputs.}
\rnewOneCN{The first stage and the second stage respectively request the first and the second input channels of the pre-fetched elements for computations, and the results of the two stages accumulate inside the SAU.}

\rnewOneCN{\Rone{Fig.~\ref{fig:mult_precision_FF_CF}~(b) and (c)} 
illustrate how to the pre-fetch and the request of 16-bit weight (W16) and 8-bit weight (W8) elements in a two-stage convolution operations. 
In W16+CF mode, we prefetch 2 weights along the output channel dimension to enhance parallel computing within a single lane, where the number of weights is determined by $TILE\_C$.}
\rnewOneCN{The elements at the same position in the first channel of both weights are simultaneously requested by the SAU and participate in the convolution computations for the first stage.}
\rnewOneCN{After the first stage, the second channel elements of the weights participate in the computations.}
\rnewOneCN{In W8+FF mode, as discussed in Sec.~\ref{subsec:overall_archi}, unified W8 element has a parallelism of 4 along the input channel dimension, thereby enhancing computational efficiency.}
\rnewOneCN{Weights are reused in the second stage to minimize off-chip data movement when the computations in the first stage are completed.}
\rnewOneCN{
In summary, the FF strategy takes advantage of calculating larger convolution kernels due to its high data reuse. 
The CF strategy is suitable for smaller convolution kernels with low reuse requirements, 
as it reduces the memory footprint of partial results and avoids additional output transfer consumption.}
\section{Experimental Results}\label{sec:res}
\subsection{ {Experimental Setup}}
\rnewOneCN{To highlight the effectiveness of our design, we select Ara\cite{Cavalcante2022NewAra} as a baseline design for comparison, which is the first open-source implementation of RISC-V vector processor.}
Several popular DNNs are implemented as benchmarks for evaluation, including VGG16~\cite{VGG16ACPR15}, ResNet18~\cite{ResNetCVPR16}, GoogLeNet~\cite{GoogleNetCVPR15}, and SqueezeNet~\cite{SqueezeNetArXiv16}.
\rnewOneCN{Note that the evaluated metric is area efficiency (GOPS/$mm^2$), measured across the convolutional layers in the DNN model using cycle-accurate simulation with QuestaSim, consistent with the experimental method in~\cite{Cavalcante2022NewAra}.}
To evaluate the impact of our architectural modifications at the hardware level, 
\rnewOneCN{we synthesize both SPEED and Ara using Synopsys Design Compiler 2014.09 on the TSMC 28 nm process.
For a fair comparison, we use 4 lanes and a vector length of 4096 bits for both SPEED and Ara.}
\rnewOneCN{And we set both $TILE\_R$ and $TILE\_C$ of SAU in each SPEED's lane to 4 for these experiments.}
Finally, we conduct a comprehensive comparison between our design and Ara to demonstrate the effectiveness of SPEED in multi-precision DNN deployments.
\vspace{-1em}
\begin{figure}[!htbp]
  \centering
    \includegraphics[width=1\columnwidth]{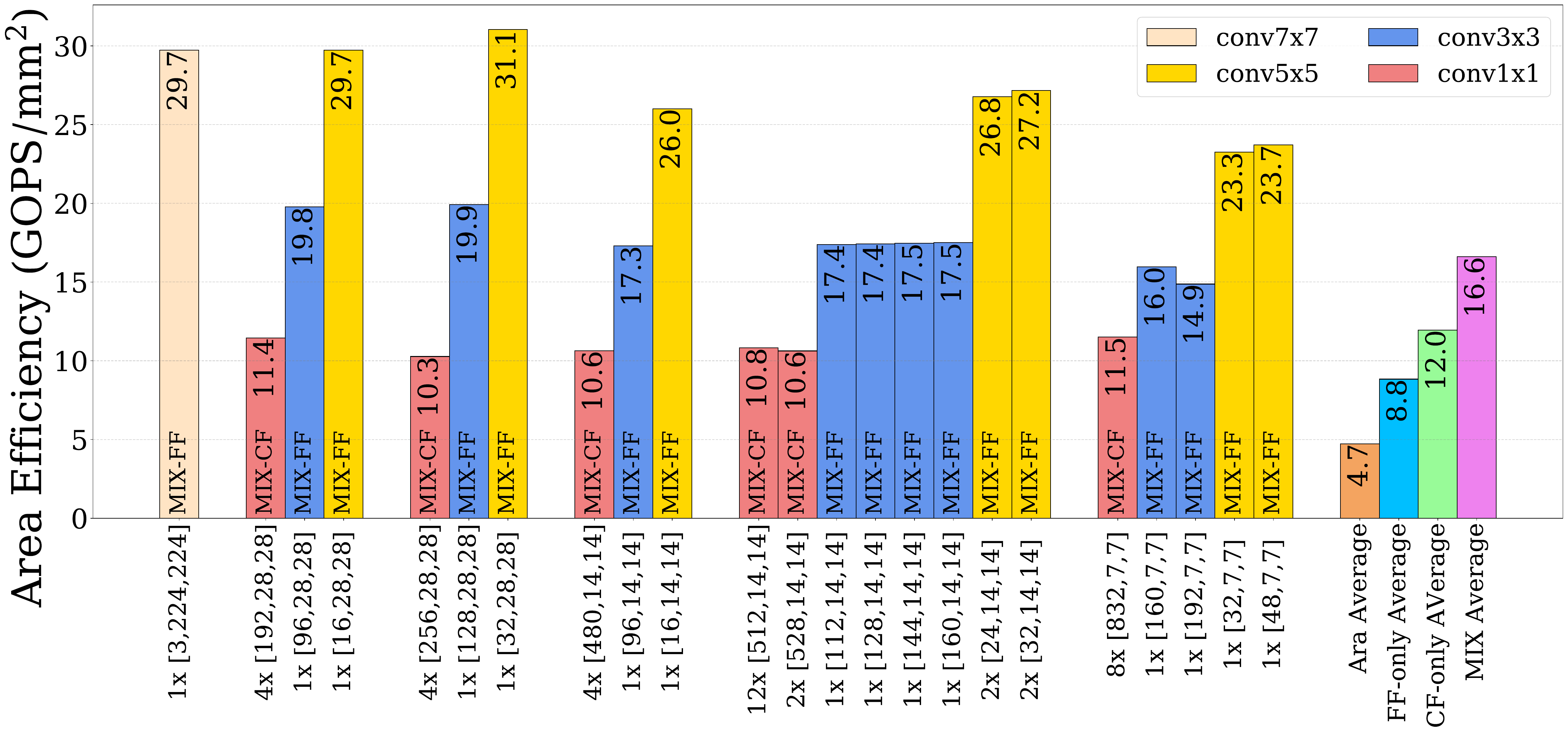}
      \caption{Layer-wise area efficiency breakdown of GoogLeNet on SPEED under 16-bit precision. Our mixed dataflow strategy surpasses the FF-only and CF-only strategies by 1.88$\times$ and 1.38$\times$, respectively.}
        \label{fig:googlenet}
    \vspace{-1.2em}
\end{figure}

\subsection{Model Evaluation}
\rnewOneCN{To evaluate the impact of convolution kernel size on different dataflow strategies, a layer-wise evaluation of GoogleNet, 
which employs diverse convolution kernel sizes, is conducted on SPEED using various strategies under the 16-bit precision condition.}  
The mixed strategy dynamically selects the FF-only or CF-only strategy with the best performance in each layer to further improve area efficiency.
\rnewOneCN{As shown in Fig.~\ref{fig:googlenet}, 
the mixed dataflow strategy achieves a area efficiency improvement of 1.88$\times$ and 1.38$\times$ in comparison with the FF-only and CF-only dataflow strategies, respectively.
Meanwhile, compared with Ara, the FF-only and CF-only dataflow strategies have a area efficiency improvement of 1.87$\times$ and 2.55$\times$, respectively, and the mixed dataflow strategy achieves a 3.53$\times$ increase in area efficiency.
To more clearly illustrate the composition of the mixed strategy in the evaluation of GoogleNet, Fig.~\ref{fig:googlenet} presents a layer-wise breakdown of the mixed strategy and annotates the specific strategy used.
The results indicate that the CF-only strategy is better suited for conv1x1, while the FF-only strategy is suitable for other convolution kernels, where convKxK denotes a convolutional operator with a kernel size of $K$.}
\rnewOneCN{Meanwhile, it shows that with larger convolution kernel sizes, the area efficiency improves.}
Therefore, employing the mixed dataflow strategy can significantly enhance SPEED's performance for DNN deployment.
\vspace{-1em}
\begin{figure}[!htbp]
  \centering
    \includegraphics[width=1\columnwidth]{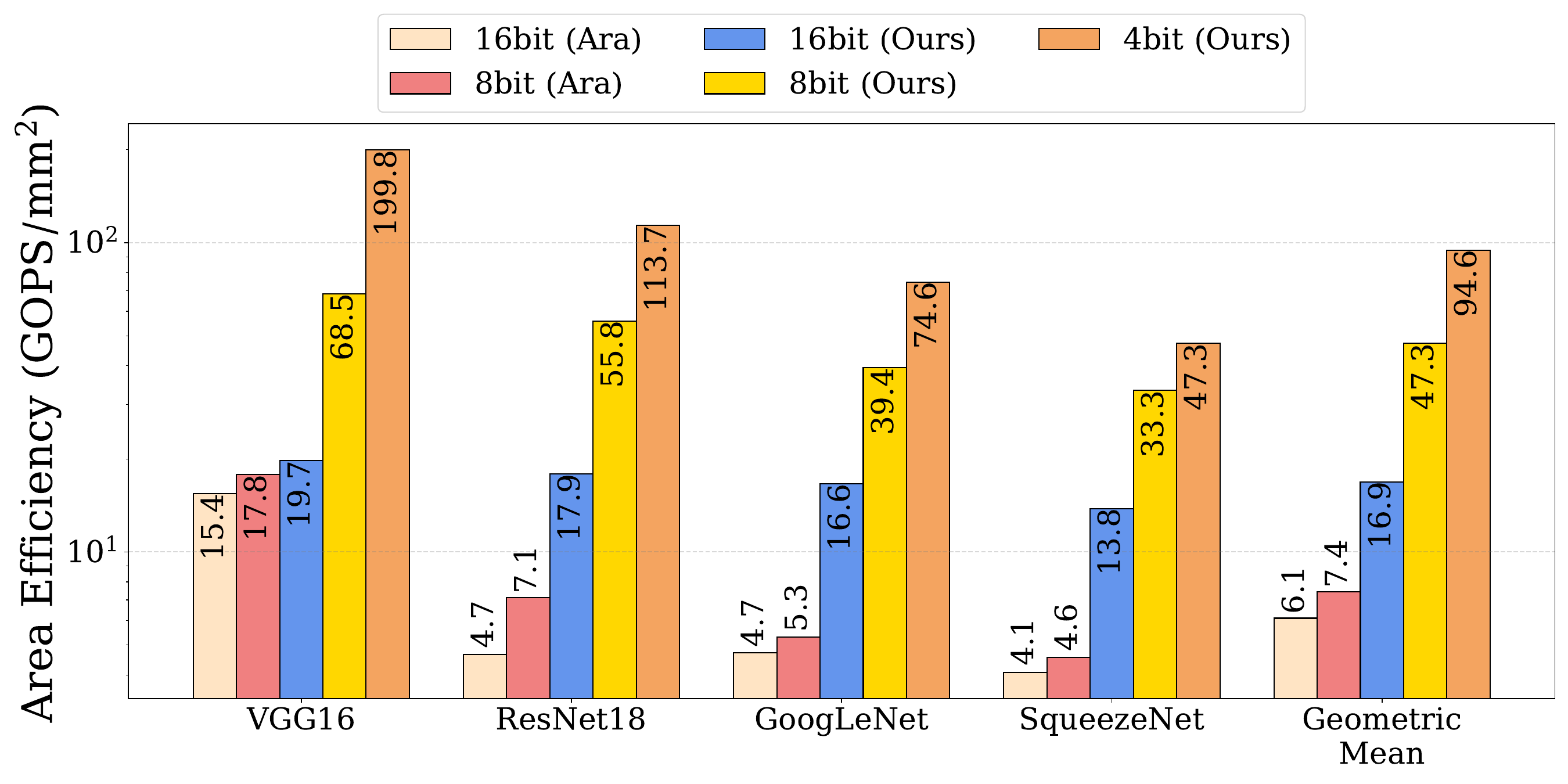}
    \caption{Average area efficiency under multi-precision DNN benchmarks, SPEED outperforms Ara by 2.77$\times$ and 6.39$\times$ at 16-bit and 8-bit precision, respectively.}
        \label{fig:per_model}
        \vspace{-0.5em}
\end{figure}

As illustrated in Fig.~\ref{fig:per_model}, we conduct a comprehensive area efficiency evaluation of SPEED with the mixed dataflow strategy using multiple popular DNNs across various precisions.
SPEED achieves 2.77$\times$ and 6.39$\times$ higher area efficiency over Ara on average under 16-bit and 8-bit precision conditions, respectively.
Moreover, SPEED enables efficient 4-bit inference with an average area efficiency of up to 94.6 GOPS/$mm^2$, surpassing the best of Ara by 12.78$\times$ on these DNN benchmarks.
\vspace{-1em}
\begin{figure}[!htp]
    \centering
    \includegraphics[width=1\columnwidth]{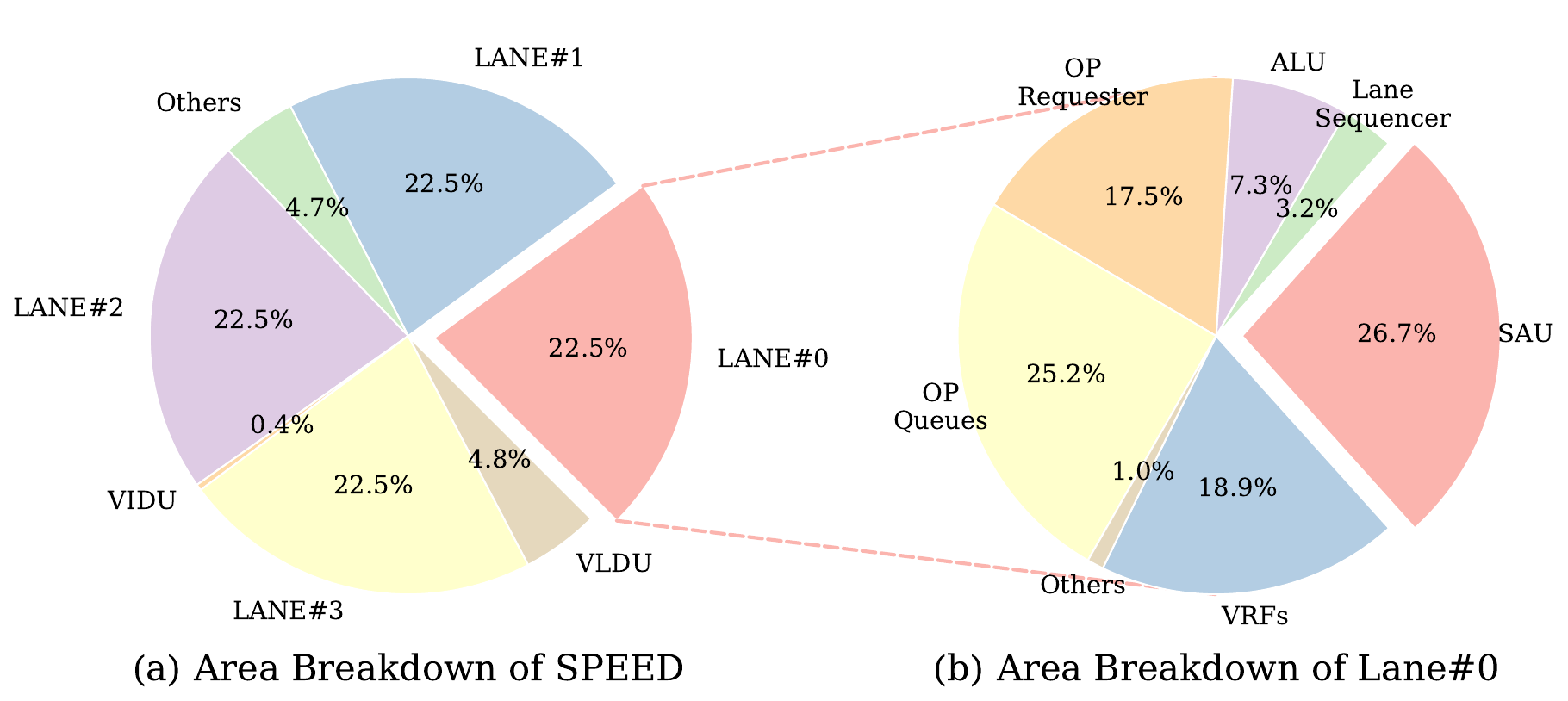}
        \caption{\rTwoFC{Area Breakdown of (a) SPEED and (b) a single lane. SAU occupies only 26\% of the area in a single lane while achieving significant computational performance.}}
    \label{fig:area_breakdown}
    \vspace{-1em}
\end{figure}
\subsection{ {Analysis of Synthesized Results}}
Fig.~\ref{fig:area_breakdown}~(a) illustrates the area breakdown of SPEED with a total area of 1.10 $mm^2$.
It reveals that the majority of SPEED's area, up to 90\%, is occupied by the lanes.
Moreover, Fig.~\ref{fig:area_breakdown}~(b) shows that the area of a lane is primarily consumed by four components: OP Queues (25\%), OP Requester (17\%), VRFs (18\%) and SAU (26\%).
Notably, SAU serves as a key component of SPEED, enabling a 12.78$\times$ area efficiency improvement in DNN inference performance on SPEED compared to Ara.
Despite the substantial performance boost, SAU accounts for only 26\% of the lane area, which corresponds to about 24\% of the total area of SPEED.
\begin{table}[!htbp]
\vspace{-1 em}
\centering
\caption{Synthesized Results of Ara and SPEED } 
\renewcommand \arraystretch{1.25} 
\label{tab:Related_Work_Summary}
\resizebox{0.42\textwidth}{!}{
\begin{tabular}{ccc}
\hline
&Ara\cite{Cavalcante2022NewAra}                       
& SPEED (Ours)                     \\ \hline
ISA                                       
& \texttt{RV64GCV1.0}  
& \texttt{RV64GCV1.0}               \\
Frequency                        
& 500 MHz @ 0.9 V
& 500 MHz @ 0.9 V         \\
Chip Area ($mm^2$)    & 0.44     & 1.10              \\
Int. Formats (bit)   & 8, 16, 32, and 64   & 4, 8, 16, 32, and 64 \\ 
Power (mW)           & 61.14                & 215.16  \\
\hline
\multirow{3}{*}{\shortstack{Peak Int. Throughput \\ (GOPS)}} 
 & 6.82 (16b)        & 34.89 (16b)            \\
& 22.95 (8b)        & 93.65 (8b)             \\
&- & 287.41 (4b)  \\
 \hline
\multirow{3}{*}{\shortstack{Peak Int. Area Efficiency \\ (GOPS/$mm^2$)}}  
& 15.51 (16b)     & 31.72 (16b)                         \\
& 52.16 (8b)      & 85.13 (8b)                         \\
& -               & 261.28 (4b)                  \\ \hline
\multirow{3}{*}{\shortstack{Peak Int. Energy Efficiency \\ (GOPS/W)}}
& 111.61 (16b)     & 162.15 (16b)                         \\
& 373.68 (8b)      & 435.25 (8b)                         \\
& -               & 1335.79 (4b)                  \\ 
 \hline
\end{tabular}
}
\vspace{-0.5em}
\end{table}

\rnewOneCN{Table~\ref{tab:Related_Work_Summary} provides a comprehensive comparison between SPEED and Ara~\cite{Cavalcante2022NewAra}, showing the peak throughput results of Ara and SPEED across various precision conditions through evaluating each convolutional layer in all DNN benchmarks.}
Based on the results in Table~\ref{tab:Related_Work_Summary}, SPEED shows 5.12$\times$ and 4.14$\times$ throughput improvement under 16-bit and 8-bit precision condition, respectively.
In addition, SPEED enables efficient 16-bit and 8-bit inference that surpasses the best of Ara by 2.04$\times$ and 1.63$\times$, respectively, on area efficiency.
In terms of energy efficiency, SPEED also demonstrates enhancements of 1.45$\times$ for 16-bit and 1.16$\times$ for 8-bit cases, respectively.

\section{Conclusion}\label{sec:concls}

In this paper, a scalable RISC-V vector processor, namely SPEED, is proposed to enable efficient multi-precision DNN inference. 
SPEED develops several customized RISC-V vector instructions to efficiently support DNN operations ranging from 4-bit to 16-bit precision.
It also enhances the parallel processing capabilities of scalable modules to boost throughput.
Moreover, a mixed dataflow strategy is presented that significantly improves the computational efficiency for various convolution kernels.
Experimental results show that SPEED achieves 2.04$\times$ and 1.63$\times$ higher area efficiency over the pioneer open-source vector processor, Ara, under 16-bit and 8-bit precision conditions, respectively, demonstrating its good potential for multi-precision DNN inference.


%
\IEEEpeerreviewmaketitle


\section*{Acknowledgment}

This work was supported in part by the National Key R\&D Program of China under Grant 2022YFB4400604, in part by the National Natural Science Foundation of China under Grant 62174084 and 62341408, and in part by the AI \& AI for Science Project of Nanjing University.



%

\clearpage
\normalem
\bibliographystyle{IEEEtran}
\bibliography{ref}	

\begin{thebibliography}{10}
\providecommand{\url}[1]{#1}
\csname url@samestyle\endcsname
\providecommand{\newblock}{\relax}
\providecommand{\bibinfo}[2]{#2}
\providecommand{\BIBentrySTDinterwordspacing}{\spaceskip=0pt\relax}
\providecommand{\BIBentryALTinterwordstretchfactor}{4}
\providecommand{\BIBentryALTinterwordspacing}{\spaceskip=\fontdimen2\font plus
\BIBentryALTinterwordstretchfactor\fontdimen3\font minus \fontdimen4\font\relax}
\providecommand{\BIBforeignlanguage}[2]{{%
\expandafter\ifx\csname l@#1\endcsname\relax
\typeout{** WARNING: IEEEtran.bst: No hyphenation pattern has been}%
\typeout{** loaded for the language `#1'. Using the pattern for}%
\typeout{** the default language instead.}%
\else
\language=\csname l@#1\endcsname
\fi
#2}}
\providecommand{\BIBdecl}{\relax}
\BIBdecl

\bibitem{xu2022towards}
Y.~Xu \emph{et~al.}, ``{Towards Developing High Performance RISC-V Processors Using Agile Methodology},'' in \emph{2022 55th IEEE/ACM International Symposium on Microarchitecture (MICRO)}, 2022, pp. 1178--1199.

\bibitem{Rossi2021VegaAT}
D.~Rossi \emph{et~al.}, ``{Vega: A Ten-Core SoC for IoT Endnodes With DNN Acceleration and Cognitive Wake-Up From MRAM-Based State-Retentive Sleep Mode},'' \emph{IEEE Journal of Solid-State Circuits (JSSC)}, vol.~57, no.~1, pp. 127--139, 2022.

\bibitem{Garofalo2023Dustin}
A.~Garofalo \emph{et~al.}, ``{Dustin: A 16-Cores Parallel Ultra-Low-Power Cluster With 2b-to-32b Fully Flexible Bit-Precision and Vector Lockstep Execution Mode},'' \emph{IEEE Transactions on Circuits and Systems I: Regular Papers (TCAS-I)}, vol.~70, no.~6, pp. 2450--2463, 2023.

\bibitem{Garofalo2023DARKSIDEAH}
A.~Garofalo \emph{et~al.}, ``{DARKSIDE: A Heterogeneous RISC-V Compute Cluster for Extreme-Edge On-Chip DNN Inference and Training},'' \emph{IEEE Open Journal of the Solid-State Circuits Society (OJSSCS)}, vol.~2, pp. 231--243, 2022.

\bibitem{yun2023TCSII}
M.~Perotti \emph{et~al.}, ``{Yun: An Open-Source, 64-Bit RISC-V-Based Vector Processor With Multi-Precision Integer and Floating-Point Support in 65-nm CMOS},'' \emph{IEEE Transactions on Circuits and Systems II: Express Briefs (TCAS-II)}, vol.~70, no.~10, pp. 3732--3736, 2023.

\bibitem{Cavalcante2022NewAra}
M.~Perotti \emph{et~al.}, ``{A “New Ara” for Vector Computing: An Open Source Highly Efficient RISC-V V 1.0 Vector Processor Design},'' in \emph{2022 IEEE 33rd International Conference on Application-specific Systems, Architectures and Processors (ASAP)}, 2022, pp. 43--51.

\bibitem{Cavalcante2020AraA1TVLSI}
M.~Cavalcante \emph{et~al.}, ``{Ara: A 1-GHz+ Scalable and Energy-Efficient RISC-V Vector Processor With Multiprecision Floating-Point Support in 22-nm FD-SOI},'' \emph{IEEE Transactions on Very Large Scale Integration Systems (TVLSI)}, vol.~28, no.~2, pp. 530--543, 2020.

\bibitem{Askarihemmat2023QuarkAI}
M.~Askarihemmat \emph{et~al.}, ``{Quark: An Integer RISC-V Vector Processor for Sub-Byte Quantized DNN Inference},'' in \emph{2023 IEEE International Symposium on Circuits and Systems (ISCAS)}, 2023, pp. 1--5.

\bibitem{Theo2023sparq}
T.~Dupuis \emph{et~al.}, ``{Sparq: A Custom RISC-V Vector Processor for Efficient Sub-Byte Quantized Inference},'' in \emph{2023 21st IEEE Interregional NEWCAS Conference (NEWCAS)}, 2023, pp. 1--5.

\bibitem{he2023agile}
Z.~He \emph{et~al.}, ``{Agile Hardware and Software Co-design for RISC-V-based Multi-precision Deep Learning Microprocessor},'' in \emph{2023 28th Asia and South Pacific Design Automation Conference (ASP-DAC)}, 2023, pp. 490--495.

\bibitem{huang2023precision}
L.~Huang \emph{et~al.}, ``{A Precision-Scalable RISC-V DNN Processor with On-Device Learning Capability at the Extreme Edge},'' \emph{arXiv preprint arXiv:2309.08186}, 2023.

\bibitem{Waterma2016EECS}
\BIBentryALTinterwordspacing
A.~Waterman, ``{Design of the RISC-V Instruction Set Architecture},'' Ph.D. dissertation, EECS Department, University of California, Berkeley, Jan 2016. [Online]. Available: \url{http://www2.eecs.berkeley.edu/Pubs/TechRpts/2016/EECS-2016-1.html}
\BIBentrySTDinterwordspacing

\bibitem{askarihemmat2023barvinn}
M.~Askarihemmat \emph{et~al.}, ``{BARVINN: Arbitrary Precision DNN Accelerator Controlled by a RISC-V CPU},'' in \emph{2023 28th Asia and South Pacific Design Automation Conference (ASP-DAC)}, 2023, pp. 483--489.

\bibitem{wu2021flexible}
X.~Wu \emph{et~al.}, ``{A Flexible and Efficient FPGA Accelerator for Various Large-scale and Lightweight CNNs},'' \emph{IEEE Transactions on Circuits and Systems I: Regular Papers (TCAS-I)}, vol.~69, no.~3, pp. 1185--1198, 2022.

\bibitem{fang2021accelerating}
C.~Fang \emph{et~al.}, ``{Accelerating 3D Convolutional Neural Networks Using 3D Fast Fourier Transform},'' in \emph{2021 IEEE International Symposium on Circuits and Systems (ISCAS)}, 2021, pp. 1--5.

\bibitem{ding2018quantized}
R.~Ding \emph{et~al.}, ``{Quantized Deep Neural Networks for Energy Efficient Hardware-based Inference},'' in \emph{2018 23rd Asia and South Pacific Design Automation Conference (ASP-DAC)}, 2018, pp. 1--8.

\bibitem{tian2023bebert}
J.~Tian \emph{et~al.}, ``{BEBERT: Efficient and Robust Binary Ensemble BERT},'' in \emph{2023 IEEE International Conference on Acoustics, Speech and Signal Processing (ICASSP)}, 2023, pp. 1--5.

\bibitem{zhou2018explicit}
A.~Zhou \emph{et~al.}, ``{Explicit Loss-Error-Aware Quantization for Low-Bit Deep Neural Networks},'' in \emph{2018 IEEE/CVF Conference on Computer Vision and Pattern Recognition (CVPR)}, 2018, pp. 9426--9435.

\bibitem{zaruba2019cost}
F.~{Zaruba} \emph{et~al.}, ``{The Cost of Application-Class Processing: Energy and Performance Analysis of a Linux-Ready 1.7-GHz 64-Bit RISC-V Core in 22-nm FDSOI Technology},'' \emph{IEEE Transactions on Very Large Scale Integration Systems (TVLSI)}, vol.~27, no.~11, pp. 2629--2640, Nov 2019.

\bibitem{Li2022APE}
K.~Li \emph{et~al.}, ``{A Precision-Scalable Energy-Efficient Bit-Split-and-Combination Vector Systolic Accelerator for NAS-Optimized DNNs on Edge},'' in \emph{2022 Design, Automation \& Test in Europe Conference \& Exhibition (DATE)}, 2022, pp. 730--735.

\bibitem{VGG16ACPR15}
S.~Liu \emph{et~al.}, ``{Very deep convolutional neural network based image classification using small training sample size},'' in \emph{2015 3rd IAPR Asian Conference on Pattern Recognition (ACPR)}, 2015, pp. 730--734.

\bibitem{ResNetCVPR16}
K.~He \emph{et~al.}, ``{Deep Residual Learning for Image Recognition},'' in \emph{2016 IEEE Conference on Computer Vision and Pattern Recognition (CVPR)}, 2016, pp. 770--778.

\bibitem{GoogleNetCVPR15}
C.~Szegedy \emph{et~al.}, ``{Going Deeper with Convolutions},'' in \emph{2015 IEEE Conference on Computer Vision and Pattern Recognition (CVPR)}, 2015, pp. 1--9.

\bibitem{SqueezeNetArXiv16}
\BIBentryALTinterwordspacing
F.~N. Iandola \emph{et~al.}, ``{SqueezeNet: AlexNet-level Accuracy with 50x Fewer Parameters and $<$0.5MB Model Size},'' \emph{CoRR}, 2016. [Online]. Available: \url{http://arxiv.org/abs/1602.07360}
\BIBentrySTDinterwordspacing

\end{thebibliography}

\end{document}